\documentclass[a4paper]{report}
\usepackage[utf8]{inputenc}
\usepackage[T1]{fontenc}
\usepackage{RJournal}
\usepackage{amsmath,amssymb,array}
\usepackage{booktabs}

\usepackage{subcaption}

\begin{document}

\sectionhead{Contributed research article}
\volume{XX}
\volnumber{YY}
\year{20ZZ}
\month{AAAA}

\begin{article}
\title{stratamatch: Prognostic Score Stratification using a Pilot Design}
\author{by Rachael C. Aikens, Joseph Rigdon, Justin Lee, Michael Baiocchi, Andrew B. Goldstone, Peter Chiu, Y. Joseph Woo, and Jonathan H. Chen}

\maketitle

\abstract{
Optimal propensity score matching has emerged as one of the most ubiquitous approaches for causal inference studies on observational data; However, outstanding critiques of the statistical properties of propensity score matching have cast doubt on the statistical efficiency of this technique, and the poor scalability of optimal matching to large data sets makes this approach inconvenient if not infeasible for sample sizes that are increasingly commonplace in modern observational data. The \CRANpkg{stratamatch} package provides implementation support and diagnostics for `stratified matching designs,' an approach which addresses both of these issues with optimal propensity score matching for large-sample observational studies. First, stratifying the data enables more computationally efficient matching of large data sets. Second, \pkg{stratamatch} implements a `pilot design' approach in order to stratify by a prognostic score, which may increase the precision of the effect estimate and increase power in sensitivity analyses of unmeasured confounding.}

\section{Introduction}\label{sec:intro}

To make causal inference from observational data, researchers must address concerns that effect estimates may be biased due to confounding factors -- baseline characteristics of the individuals in the study that influence both their selection of treatment and their probable outcome. Matching methods seek to account for this self-selection by grouping treated and control individuals with similar baseline characteristics. One of the most common such methods, propensity score matching, pairs individuals who appear to have had similar probabilities of receiving the treatment according to their baseline characteristics \citep{rosenbaum1983central}, with the goal of coercing the data set into a form that resembles a fully-randomized controlled trial \citep{king2016propensity, rosenbaum2010designbook, hernan2016using}. However, propensity score matching can only address bias to due \textit{measured} baseline covariates, necessitating sensitivity analyses to interrogate the potential of bias due to \textit{unmeasured} confounding \citep{rosenbaum2005sensitivity, rosenbaum2010designbook}.

In their provocative article ``Why Propensity Should Not Be Used for Matching,'' King and Nielson argue that that the fully randomized controlled trial -- the design emulated by propensity score matching -- is less statistically efficient than the block-randomized controlled experiment \citep{king2016propensity}. In block-randomized designs, individuals are stratified by prognostically important covariates (e.g., for a clinical trial: sex, age group, smoking status) \textit{prior} to randomization in order to reduce the heterogeneity between the treatment and control groups.  In the experimental context, these efforts to reduce heterogeneity between compared groups help to increase the precision of the treatment effect estimate. In observational settings, reducing this type of heterogeneity not only improves precision but increases the robustness of the study's conclusions to being explained away by the possibility of unobserved confounding \citep{rosenbaum2005heterogeneity, aikens2020pilot}.  The stratified matching design -- in which observations are stratified prior to matching within strata -- attempts to emulate the block-randomized controlled trial design in the observational context in order to secure these statistical benefits over pure propensity score matching. In addition, since the computation required for optimal matching can be quite time-consuming for studies of more than a few thousand observations, stratified matching designs could greatly improve the scalability of optimal matching.  While the computational complexity of optimal matching is non-polynomial, the process of matching a stratified data set with a constant stratum size scales much more favorably with sample size (see section \ref{subsec:advanced_matching}).

While a variety of packages exist for matching subjects in observational studies, limited support exists for researchers seeking to implement a stratified matching design.  The popular \CRANpkg{matchit} package \citep{ho2011matchit} is a user-friendly option for common propensity score matching designs and related approaches, \CRANpkg{optmatch} \citep{hansen2006optmatch} and \CRANpkg{DOS2} \citep{rosenbaum2019DOS2} are a powerful combination for implementing a variety of more complicated optimal matching schemes, and \CRANpkg{nearfar} \citep{rigdon2018nearfar} implements a different form of matching for the instrumental variable study. The primary goal of \pkg{stratamatch} is to make stratified matching and prognostic score designs accessible to a wider variety of applied researchers, and to suggest a suite of diagnostic tools for the stratified observational study. In favorable settings, these designs could not only increase the precision and robustness of inference but could facilitate optimal matching of sample sizes for which this technique was previously computationally impractical. 

This paper discusses the methodological contributions of \pkg{stratamatch} -- in particular the implementation of a novel pilot design approach suggested by \citet{aikens2020pilot} (section \ref{sec:Study_design}) -- and summarizes the package implementation (Section \ref{sec:software}) with illustrative examples (Section \ref{sec:illustrations}).  While \pkg{stratamatch} may substantially improve the scalability of optimal matching for some large data sets, the main objective of the package is not to implement a computationally complex task but to make sophisticated study design tools and concepts accessible to a wide variety of researchers. 

\section{Study design}\label{sec:Study_design}

\subsection[Pilot designs]{A prognostic score stratification pilot design}\label{subsec:pilot_designs}

Stratifying a data set based on baseline variation prior to matching reduces the heterogeneity between matched sets with respect to that baseline variation. But what baseline characteristics should be used? One option is to select prognostically important covariates by hand, based on expert knowledge. However, in practice, this ``manual'' stratification process often produces strata that vary wildly in size and composition. Some strata may be so small or so imbalanced in their composition of treated and control individuals that it is difficult to find high-quality matches or many observations are thrown away.  Other strata may be so large that matching within them is still computationally infeasible.

The \code{auto\_stratify} function in \pkg{stratamatch} divides subjects into strata using a \dfn{prognostic score} (see \citet{hansen2008prognostic}), which summarizes the baseline variation most important to the outcome.  In addition to producing strata of more regular size and composition, balancing treatment and control groups based on the prognostic score may confer several statistical benefits: increasing precision \citep{aikens2020pilot, leacy2014joint}, providing some protection against mis-specification of the propensity score \citep{leacy2014joint, antonelli2018doubly}, and decreasing the susceptibility of an observed effect to being explained away by unobserved confounding \citep{rosenbaum1983central, aikens2020pilot}.  However, fitting the prognostic score on the same data set raises concerns of overfitting and may lead to biased effect estimates \citep{hansen2008prognostic, abadie2018endogenous}.  For this reason, \citep{aikens2020pilot} suggest using a \dfn{pilot design} for estimating the prognostic score.

Central to the pilot design concept is maintaining separation between the design and analysis phases of a study (see table \ref{tab:defs}, or for more information \citet{goodman2017designthinking} and \citet{rubin2008design}). Using an observational pilot design, the researchers partition their data set into an \dfn{analysis set} and a held-aside \dfn{pilot set}. Outcome information in the pilot set can be observed (e.g. to fit a prognostic score), and the information gained can be used to inform the study design. Subsequently, in order to preserve the separation of the study design from the study analysis, the individuals from the pilot set are omitted from the main analysis (i.e., they are not reused in the analysis set). The primary insight of the pilot design is that reserving all of the observations in a study for the analysis phase (i.e., in the analysis set) is not always better.  Rather, clever use of data in the design phase (i.e., in the pilot set) may facilitate the design of stronger studies.

In the \pkg{stratamatch} approach, a random subsample of controls is extracted as a pilot set to fit a prognostic model, and that model is then used to estimate prognostic scores on the mix of control and treated individuals in the analysis set.  The observations in the analysis set can then be stratified based on the quantiles of the estimated prognostic score, and matched by propensity score within strata (see section \ref{sec:software}).

\begin{table}[t!]
\centering
 \begin{tabular} {p{2.8cm}p{10.3cm}} \toprule
 Term      &  Description \\ \midrule
\textbf{Design phase} & Phase of a study in which the researcher considers what kinds of data will provide the strongest information to address the question at hand (e.g., randomization, sampling, matching, inverse probability weighting). The goal of the design phase is to obtain data which will provide strong inference. \\
\textbf{Analysis phase} & Phase of a study in which the data that comes from the design phase are summarized into statistics. Inference and sensitivity analyses are performed. \\
\textbf{Pilot Design} & An observational study approach in which some data is spent in the design phase to improve the study design/preprocessing. \\
\textbf{Pilot Set} & A subset of data extracted to be used in the design phase. \\
\textbf{Analysis set} & The set of data reserved for inference in the analysis phase. \\
\textbf{Propensity score} & Probability of assignment to the treatment group based on measured baseline characteristics. \\
\textbf{Prognostic score} & Expectation of the outcome in the absence of treatment based on measured baseline characteristics. \\
\textbf{Prognostic model} & A model (e.g. logistic regression) used to estimate prognostic scores.\\
\textbf{Stratum} & A subset of observations in the analysis set to be matched together.\\
 \bottomrule
 \end{tabular}
 \caption{\label{tab:defs} Summary of relevant methodological terms as they apply to \pkg{stratamatch}.}
 \end{table}
 
\subsection{When to use this approach}\label{subsec:when_to_use}

\citet{aikens2020pilot} describe the scenarios in which a prognostic score matching pilot design is most useful.  Briefly, the \pkg{stratamatch} approach is best for large data sets (i.e., thousands to millions of observations), especially when the number of control observations is plentiful. This technique may be particularly useful when modeling a prognostic score with the measured covariates is straightforward, and when propensity score alone is likely to exclude certain aspects of variation highly associated with outcome but unassociated with treatment assignment. While computational gains vary, stratification tends to noticeably accelerate matching for sample sizes of 5,000 or more (see section \ref{subsec:advanced_matching}). 

Conversely, this technique is not recommended for small data sets in which each control observation is precious, especially when prognostic scores are likely to be difficult to estimate from the measured covariates (see \citet{aikens2020pilot} for a lengthy discussion).  Ideally, there should be ample control observations available to fit a usable prognostic model and still leave sufficient controls remaining to select high-quality matches for the treated individuals in the data set.  While some \pkg{stratamatch} designs may be useful for the estimation of other causal estimands, the statistical properties of prognostic pilot designs for estimands other than the average treatment effect among the treated have not yet been characterized \citep{aikens2020pilot}.

\section{Software}\label{sec:software}

The \pkg{stratamatch} function, \code{auto\_stratify}, implements the prognostic score stratification in the pilot design described above. The most basic procedure does the following:

\begin{enumerate}
    \item Partition the data set into a pilot data set and an analysis data set
    \item Fit a model for the prognostic score from the observations in the pilot set
    \item Estimate prognostic scores for the analysis set using the prognostic model
    \item Stratify the analysis set based on prognostic score quantiles.
\end{enumerate}

A call to \code{auto\_stratify} produces an \code{auto\_strata} object, which contains the analysis set, the pilot set, and other information about the strata and prognostic scores. The \pkg{stratamatch} package implements a set of diagnostic plots and tables that can be used to assess the quality of a stratification.  Example code, output, and diagnostics are provided in section \ref{sec:illustrations}. If the strata are satisfactory, the treatment and control individuals within each stratum can then be matched. The \code{strata\_match} function implements $1:k$ propensity score matching within each stratum. However, if something more complex is desired, the researcher may also select and implement their own matching scheme (see section \ref{subsec:advanced_matching}).

\section{Illustrations} \label{sec:illustrations}

\subsection{Simulated example} \label{subsec:simulated_eg}

This section demonstrates the basic functionality of \pkg{stratamatch} in simulated example.  The \code{make\_sample\_data} function generates a simple simulated data set so that users can explore the design options implemented by \pkg{stratamatch}. Below, we generate a sample of 10,000 observations and print the first few rows as an illustration.

\begin{example}
library("stratamatch")
library("dplyr")
dat <- make_sample_data(n = 10000)
head(dat)
\end{example}
\begin{example}
           X1        X2 B1 B2 C1 treat outcome
1  0.93332697 1.0728339  1  0  a     1       0
2 -0.52503178 0.3449057  1  1  b     0       1
3  1.81443979 1.0361942  1  1  a     0       0
4  0.08304562 0.3017060  1  1  a     0       1
5  0.39571880 0.5397257  0  0  c     0       0
6 -2.19366962 1.4523274  1  1  b     0       1  
\end{example}

The user should suppose that the rows of \code{dat} are individuals in an observational study, and the objective of the study is to estimate the effect of a binary treatment assignment (\code{treat}) on a binary outcome (\code{outcome}).    Columns 1-5 represent three types of measured baseline covariates: continuous (\code{X1} and \code{X2}), binary (\code{B1} and \code{B2}) and categorical (\code{C1}). For this example, we assume strongly ignorable treatment assignment - that is, roughly, there are no unmeasured confounding factors \citep{rosenbaum1983central}. (For sensitivity analyses for this assumption see, for example \citet{rosenbaum2005sensitivity}).

\subsubsection{Automatic stratification}\label{subsubsec:simulated_astrat}

The command below uses \code{auto\_stratify} to (1) partition 10\% of the controls in \code{dat} into the pilot set (2) fit a prognostic score model for \code{outcome} based on \code{X1} and \code{X2}, (3) estimate prognostic scores on the analysis set, and (4) return to us the analysis set, divided into strata of approximately 500 individuals, based on prognostic score quantiles. All of these steps are completed automatically with this function call, and the results are returned to us as \code{a.strat}.

\begin{example}
a.strat <- auto_stratify(dat, treat = "treat", prognosis = outcome ~ X1 + X2,
+    pilot_fraction = 0.1, size = 500)
\end{example}
\begin{example}
Constructing a pilot set by subsampling 10
Fitting prognostic model via logistic regression: outcome ~ X1 + X2
\end{example}

The result returned by \code{auto\_stratify} is an \code{auto\_strata} object.  Running \code{print} on this object supplies basic information about how the stratification process has been completed.

\begin{example}
print(a.strat)
\end{example}
\begin{example}
auto_strata object from package stratamatch.

Function call:
auto_stratify(data = dat, treat = "treat", prognosis = outcome ~ 
    X1 + X2, size = 500, pilot_fraction = 0.1)

Analysis set dimensions: 9234 X 8

Pilot set dimensions: 766 X 7

Prognostic Score Formula:
outcome ~ X1 + X2
\end{example}

Here, \code{auto\_stratify} has partitioned away a pilot set of 766 control individuals to fit our desired prognostic model, leaving 9,234 individuals in the analysis set.  Using the prognostic model, prognostic scores were estimated on the individuals in the analysis set, and these individuals were divided into strata with a target size of 500.  In order to record these stratification assignments, an eighth column, \code{stratum}, has been appended to the analysis set.  The number strata and range of strata sizes can be obtained from \code{summary(a.strat)}.

The analysis set and pilot set are accessible with \code{a.strat} via \code{a.strat\$analysis\_set} and \code{a.strat\$pilot\_set}, respectively.  The \code{strata\_table} (accessed via \code{a.strat\$strata\_table}) reports the strata sizes and the prognostic score quantile bins which define each stratum.

\subsubsection{Diagnostics} \label{subsubsec:simulated_diagnostics}

A major focus of the \pkg{stratamatch} package is suggesting diagnostics for the quality of stratification in observational studies.  The \code{issue\_table} reports the total size and composition of each stratum:

\begin{example}
a.strat$issue_table
\end{example}
\begin{example}
# A tibble: 19 x 6
   Stratum Treat Control Total Control_Proportion Potential_Issues          
     <int> <int>   <int> <int>              <dbl> <chr>                     
 1       1   167     319   486              0.656 none                      
 2       2   149     337   486              0.693 none                      
 3       3   160     326   486              0.671 none                      
 4       4   132     354   486              0.728 none                      
 5       5   123     363   486              0.747 none                      
 6       6   122     364   486              0.749 none                      
 7       7   146     340   486              0.700 none                      
 8       8   109     377   486              0.776 none                      
 9       9   131     355   486              0.730 none                      
10      10   132     354   486              0.728 none                      
11      11   111     375   486              0.772 none                      
12      12   108     378   486              0.778 none                      
13      13   112     374   486              0.770 none                      
14      14   122     364   486              0.749 none                      
15      15   100     386   486              0.794 none                      
16      16   109     377   486              0.776 none                      
17      17   114     372   486              0.765 none                      
18      18   107     379   486              0.780 none                      
19      19    85     401   486              0.825 Not enough treated samples
\end{example}

The \samp{Potential\_Issues} column is meant to quickly flag strata which may be problematically large, small, or imbalanced in the ratio of treated and control samples.  The \code{"Not enough treated samples"} flag for stratum 19 indicates that the proportion of treated individuals is 0.2 or lower\footnote{Note that the specific thresholds defining the potential issue flags (e.g. 20\% treated individuals or fewer) are not universal cutoffs but guidelines meant to draw researchers' attention to possible irregularities.}.  This is a relatively common issue, which is often easily addressed (see section \ref{sec:troubleshooting}).

\begin{figure}[htbp]
\centering
    \begin{subfigure}[t]{2.25in}
        \centering
        \includegraphics[width = 2.25in]{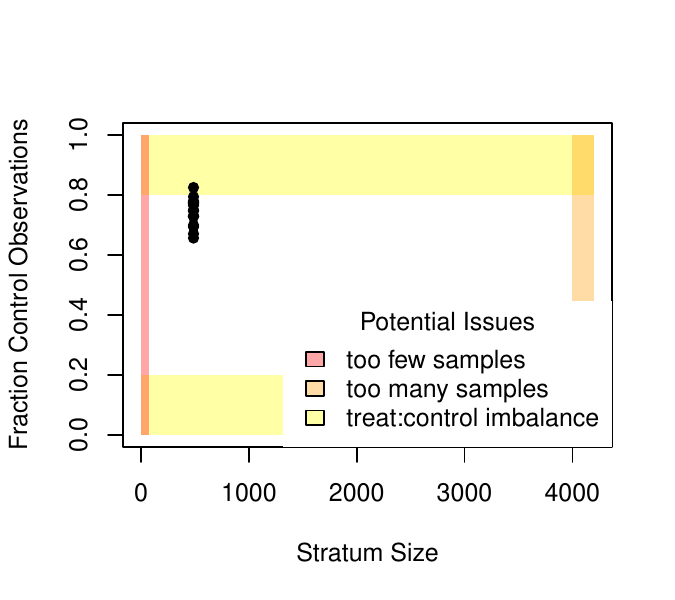}
        \caption{A size ratio plot}
    \end{subfigure}
    \begin{subfigure}[t]{2.25in}
        \centering
        \includegraphics[width = 2.25in]{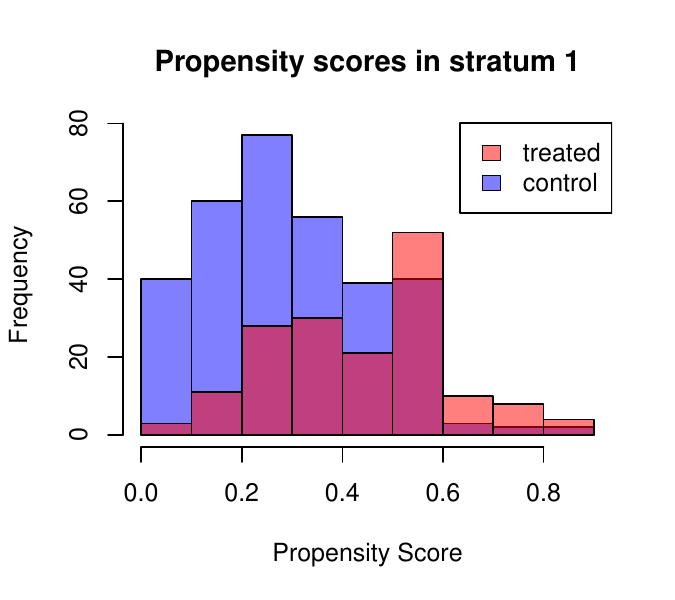}
        \caption{A propensity score histogram}
    \end{subfigure}
    \caption{ (A) A size-ratio plot, with each point representing a stratum. Yellow regions: treated to control ratio is imbalanced. Orange: strata size is large enough that matching may be computationally time-consuming. Red: strata are small enough that match quality may be poor. In a perfectly ideal stratification, all strata would fall within the white rectangle.  In practice, some stratification issues are common and easily addressed, see section \ref{sec:troubleshooting}.(B) A histogram of estimated propensity scores for a selected stratum. In an ideal scenario, there is ample overlap between treated and control individuals within each stratum. }
    \label{fig:SR_simulated}
\end{figure}

\begin{figure}[htbp]
\centering
    \begin{subfigure}[htbp]{2.5in}
        \centering
        \includegraphics[width = 2.5in]{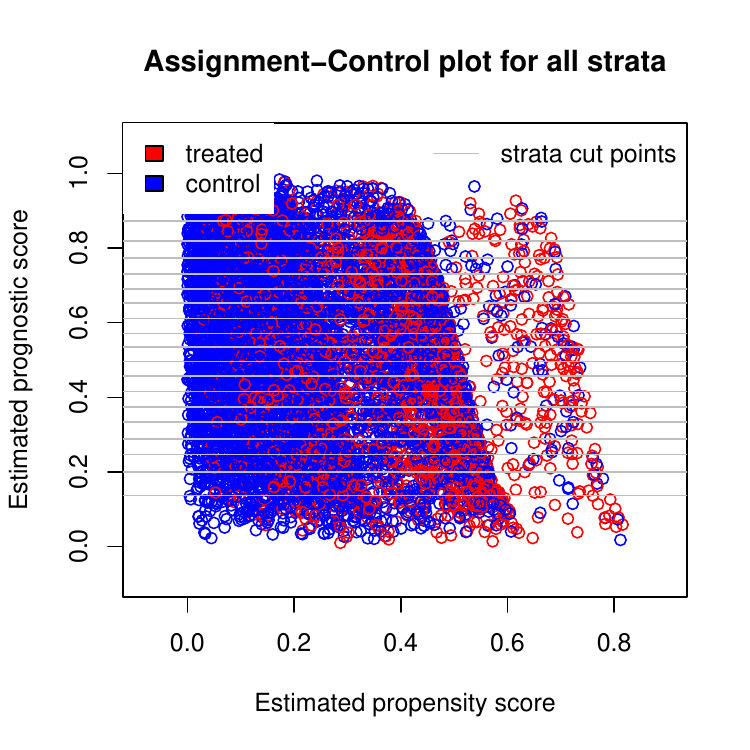}
        \caption{An assignment-control plot across all strata}
    \end{subfigure}
    \begin{subfigure}[htbp]{2.5in}
        \centering
        \includegraphics[width = 2.5in]{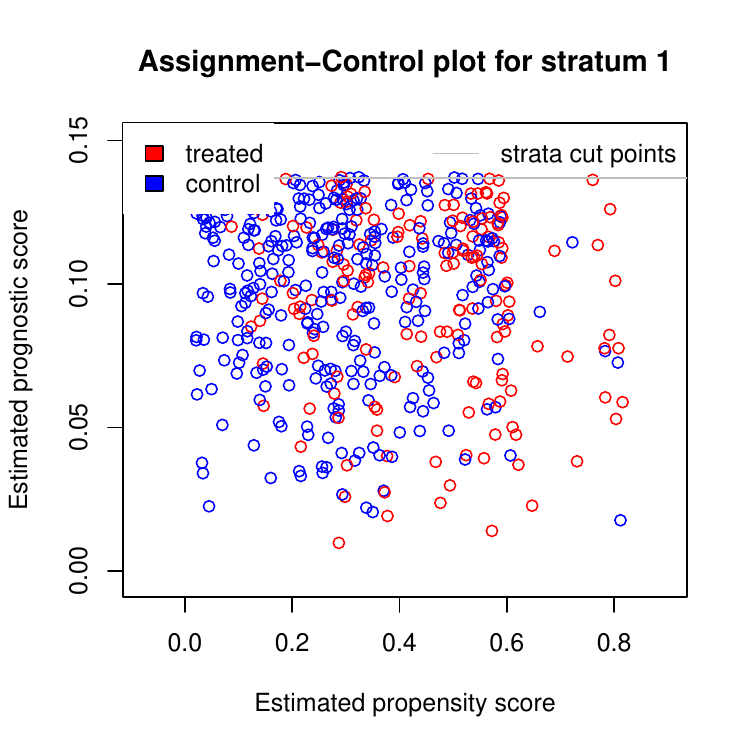}
        \caption{An assignment-control plot for stratum 1}
    \end{subfigure}
    \caption{\label{fig:AC_simulated} Assignment-control plots \citep{aikens2020pilot} showing estimated propensity score versus estimated prognostic score for each subject in the analysis set (A) or a selected stratum (B).  In an ideal scenario, there is ample overlap between treated and control individuals in terms of both prognosis and propensity (for other cases, see section \ref{sec:troubleshooting}).  Grey lines denote the prognostic score thresholds defining the strata.}
\end{figure}

The \pkg{stratamatch} package implements four diagnostic plotting options:
\begin{enumerate}
    \item \textbf{Size-Ratio Plot:} (Figure \ref{fig:SR_simulated}) Displays each stratum in the analysis set based on its size and the percentage of control observations in order to identify potentially problematic strata.
    \item \textbf{Propensity Score Histogram:} (Figure \ref{fig:SR_simulated}) Displays the distribution of estimated propensity scores across the treatment and control groups, within a single stratum or the entire analysis set. These plots are used for assessing propensity score overlap.
    \item \textbf{Assignment-Control Plot:} (Figure \ref{fig:AC_simulated}) Displays each individual based on estimated propensity score and estimated prognostic score, based on visualizations from \citet{aikens2020pilot}.  As above, these plots can display a single stratum or the entire analysis set. Assignment-control plots are useful for checking the overlap and correlation of prognostic and propensity scores.
    \item \textbf{Residual Plots:} (Not shown) Show the diagnostic plots for the prognostic model used to estimate the prognostic scores. It is essentially a wrapper for \code{plot.lm} (see the documentation for \code{plot.lm} in the base R package, \pkg{stats}). Note that since the pilot set alone is used to fit the prognostic model, only the pilot set is used for these diagnostic plots.
\end{enumerate}

The code below makes each of the plot types listed above, including two assignment-control plots: one for the entire analysis set and one for a single stratum.  The results are shown in figures \ref{fig:SR_simulated} and \ref{fig:AC_simulated}, with interpretation in the figure captions. For propensity score histograms and assignment-control plots, the \samp{propensity} argument is required, specifying how the propensity scores should be estimated. Below, the propensity score is fit on the analysis set based on a regression of treatment assignment on \samp{X1}, \samp{X2}, \samp{B1}, and \samp{B2} (for other input options, run \code{help(plot.strata)}).

\begin{example}
plot(a.strat, type = "SR")
plot(a.strat, type = "hist", propensity = treat ~ X2 + X1 + B1 + B2, stratum = 1)
plot(a.strat, type = "AC", propensity = treat ~ X2 + X1 + B1 + B2)
plot(a.strat, type = "AC", propensity = treat ~ X2 + X1 + B1 + B2, stratum = 1)
plot(a.strat, type = "residual")
\end{example}

In this example, the command \code{a.strat\$prognostic\_model} would supply the prognostic model (an \code{lm} or \code{glm} object) for further diagnostics (e.g. with \code{summary(a.strat\$prognostic\_model)}). Assessment of the prognostic model can indicate whether a sufficient number of observations has been partitioned into the pilot set (see section \ref{sec:troubleshooting}).  However, one benefit of a stratified matching design is that even an imperfect prognostic model may yield robust inference if the resulting strata are of sufficient quality to allow for a strong propensity match (see, for example theory on stratified sampling \citep{lohr2019sampling} or commentary on doubly robust matching \citep{leacy2014joint, antonelli2018doubly})

\subsubsection{Matching} \label{subsubsec:simulated_matching}

Once the data have been stratified, the user can optimally match individuals within each stratum. A 1-to-1 or 1-to-$k$ optimal propensity score matching within strata can be done automatically with the \code{strata\_match} function, as in the example below. Propensity score (the default uses all columns besides the treatment, outcome, and stratum).  This function makes essential use of the \pkg{optmatch} package (\cite{hansen2006optmatch}; see also \cite{RELAX1988}) to perform the matching within strata.

\begin{example}
mymatch <- strata_match(a.strat,  k = 1, propensity = treat ~ X1 + X2 + B1 + B2)
\end{example}
\begin{example}
Fitting propensity model: treat ~ X1 + X2 + B1 + B2 + strata(stratum)
\end{example}

The result is an optimal 1 to 1 matching within prognostic score strata. Above, \code{mymatch} is an \code{optmatch} class object, as described by the \pkg{optmatch} package \citep{hansen2006optmatch}.  For the most part, \code{mymatch} can be treated as a factor giving match assignments for each row of the data set. The command \code{summary(mymatch)} would display the number of pairs, the number of unmatched individuals, and effective sample size. For suggestions on more complex matching schemes for stratified data, see \ref{subsec:advanced_matching}.

\subsection{Real-data example: Life sustaining treatments for critical care patients}\label{subsec:real_eg}

As an applied example, the \pkg{stratamatch} package contains a re-processed version of deidentified medical data from \citet{chavez2018reversals}.  Briefly, the authors extracted demographic information, common laboratory test results, comorbidity information, and treatment team assignments for 10,157 ICU patients from the Stanford University Hospital who met their inclusion criteria. During their stay, each patient's critical care preferences are summarized with a code status.  The default -- Full Code status -- indicates no limitations on resuscitative measures, while other codes (e.g. `Do not resuscitate', or `DNR') indicate different limitations on the intensity and type of resuscitation the patient should receive if they become pulseless or apneic (i.e., their heart stops or they stop breathing). This code status is a product of complex dynamics between patient and provider.  When a patient's code status does not reflect their goals of care, patients may have life sustaining care inappropriately withheld, or they may receive aggressive treatment which does not effectively increase their quality or quantity of remaining life.

In this example, suppose a researcher wants to study whether comparable patients under the care of surgical teams vs. non-surgical teams are more likely to have their code status set to limit resuscitation (i.e., any form of `DNR'). From this we could infer tendencies that different treatment teams have in counseling and decision-making about life-sustaining treatments for the critically ill. However, the patient groups seen by surgical vs. non-surgical teams are necessarily different, because patients are assigned to treatment teams based on their reason for being in the hospital and their treatment history. Thus, a naive comparison of DNR order frequency between care team types would be misleading. To better account for these potential differences, we employ a stratified pilot matching design to compare ``treated'' (assigned to a surgical care team) individuals with ``control'' (assigned to a non-surgical care team) ones which are similar in terms of their prognostic and propensity scores.

\subsubsection{Automatic stratification}\label{subsubsec:real_astrat}

Patients must be first stratified by a prognostic score (i.e., their estimated probability of receiving a DNR order if they are not assigned to a surgical care team), before being matched on a propensity score (i.e., their estimated probability of assignment to a surgical care team). In the example below, we use  \code{auto\_stratify} on the \code{ICU\_data} to (1) partition 10\% of controls into a pilot set, (2) build a prognostic score model on that pilot set based on age (\samp{Birth.preTimeDays}), sex, and race/ethnicity (3) estimate prognostic scores on the analysis set and (4) return a stratified data set with approximately 500 individuals per stratum.

\begin{example}
ICU_astrat <- auto_stratify(data = ICU_data, treat = "surgicalTeam",
    prognosis = DNR ~ Birth.preTimeDays + Female.pre + RaceAsian.pre + 
      RaceUnknown.pre + RaceOther.pre + RacePacificIslander.pre +
      RaceBlack.pre + RaceNativeAmerican.pre + all_latinos,
    pilot_fraction = 0.1, size = 500)  
\end{example}
\begin{example}
Constructing a pilot set by subsampling 10
Fitting prognostic model via logistic regression: DNR ~ Birth.preTimeDays +
    Female.pre + RaceAsian.pre + RaceUnknown.pre +  RaceOther.pre + 
    RaceBlack.pre + RacePacificIslander.pre + RaceNativeAmerican.pre + 
    all_latinos
\end{example}

Next, we print the results.

\begin{example}
print(ICU_astrat)  
\end{example}
\begin{example}
auto_strata object from package stratamatch.

Function call:
auto_stratify(data = ICU_data, treat = "surgicalTeam", 
    prognosis = DNR ~ Birth.preTimeDays + Female.pre + RaceAsian.pre + 
        RaceUnknown.pre + RaceOther.pre + RaceBlack.pre +
        RacePacificIslander.pre + RaceNativeAmerican.pre + all_latinos,
        size = 500, pilot_fraction = 0.1)

Analysis set dimensions: 9364 X 14

Pilot set dimensions: 793 X 13

Prognostic Score Formula:
DNR ~ Birth.preTimeDays + Female.pre + RaceAsian.pre + RaceUnknown.pre + 
    RaceOther.pre + RaceBlack.pre + RacePacificIslander.pre + 
    RaceNativeAmerican.pre + all_latinos  
\end{example}
\begin{example}
summary(ICU_astrat) 
\end{example}
\begin{example}
Number of strata: 19 

	Min size: 492 	Max size: 494

Strata with Potential Issues: 2
\end{example}

We see here that \code{auto\_stratify} partitioned the data into a pilot set of 793 ``controls'' (i.e., patients not assigned to a surgical treatment team) and an analysis set of the 9,364 remaining individuals. The prognostic model was fit on the pilot set according to the formula we provided, regressing DNR code assignment on age, sex, and race. This model was used to estimate the prognostic score (probability of DNR code assignment based on demographics) for each of the 9,364 individuals in the analysis set. Finally, each individual in the analysis set was assigned to a stratum based on this score. 19 strata, each containing between 492 and 494 patients, were created.  This stratum assignment information was appended to the analysis set by adding a new 14th column, \code{stratum}.

\subsubsection{Manual stratification}\label{subsubsec:real_mstrat}

Rather than using a pilot design to build a prognostic score,  researchers may wish to stratify the data set based on discrete covariates (e.g., chosen by a domain expert). The  \code{manual\_stratify} function supports these study designs. For example, the code below bins the 10,157 patients in the data set purely based on race/ethnicity and sex. In contrast, the size-ratio plots for the automatic stratification show a much smaller range of sizes and control proportions, with fewer -- and more easily addressed -- potential issues.

\begin{example}
ICU_mstrat <- manual_stratify(data = ICU_data,
    strata_formula = surgicalTeam ~ Female.pre + RaceAsian.pre +
      RaceUnknown.pre + RaceOther.pre + RaceBlack.pre + 
      RacePacificIslander.pre + RaceNativeAmerican.pre + all_latinos) 
summary(ICU_mstrat)
\end{example}
\begin{example}
Number of strata: 16 

	Min size: 17 	Max size: 3314

Strata with Potential Issues: 9 
\end{example}

The resulting \code{manual\_strata} object has many of the same properties as an \code{auto\_strata} object from \code{auto\_stratify} and can be matched in the same way with \code{strata\_match}. However \code{manual\_strata} objects do not have a pilot set prognostic score information, and accordingly assignment-control and residual plots are not supported for these inputs.  

This more traditional manual approach may be preferred in some cases for its simplicity, and because it obviates the need to sacrifice observations to fit a prognostic model. However, selecting a binning scheme which results in favorable strata may be a time-consuming iterative process, as highlighted by the diagnostics in the following section. These issues underscore the potential usefulness of the prognostic score stratification implemented by \code{auto\_stratify}.

\subsubsection{Diagnostics}\label{subsubsec:real_diagnostics}

\begin{figure}[t!]
\centering
    \begin{subfigure}[t]{2.5in}
        \centering
        \includegraphics[width = 2.5in]{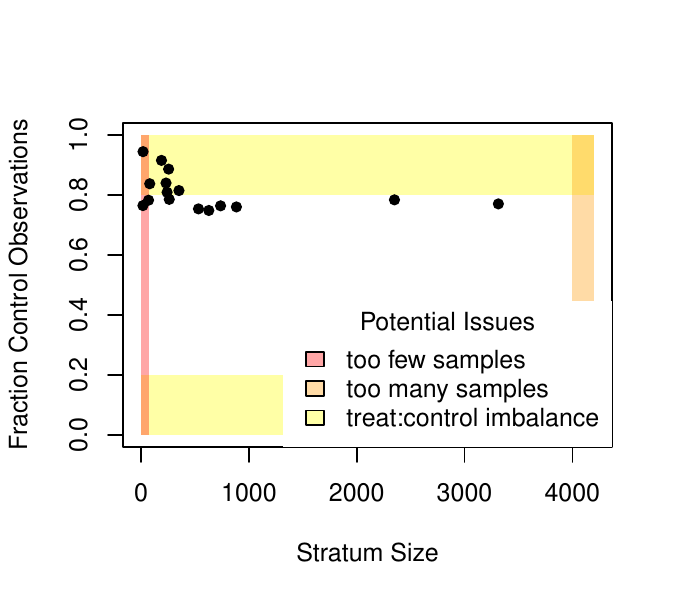}
        \caption{Size-ratio plot for manual stratification}
    \end{subfigure}
    \begin{subfigure}[t]{2.5in}
        \centering
        \includegraphics[width = 2.5in]{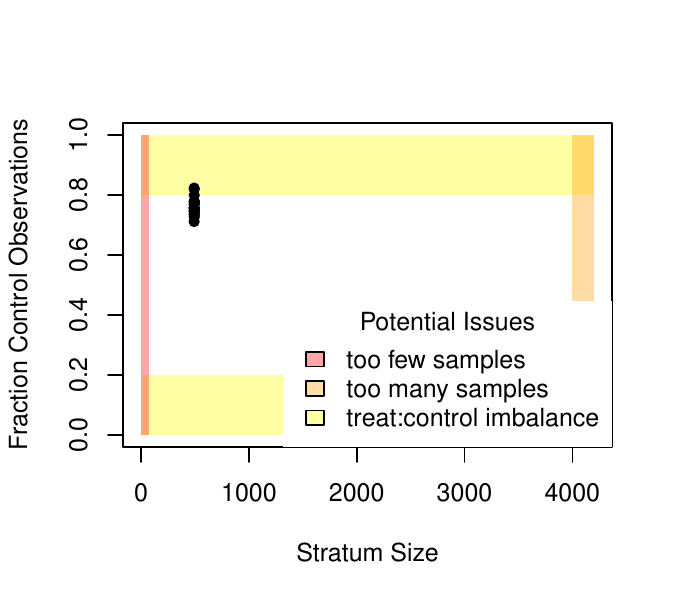}
        \caption{Size-ratio plot for automatic stratification}
    \end{subfigure}
\caption{ Size-ratio plots for (A) manual stratification on sex and race/ethnicity and (B) automatic stratifications of the same data set of ICU patients. Manual stratification often results in highly variable size and treat:control balance between strata, as reflected by the number of strata points in the shaded zones.}
\label{fig:SR_applied}
\end{figure}

Size-ratio plots for the manual and automatic stratification illustrate a common issue with manual stratification: it is often difficult to select discrete covariates which result in appropriately sized and balanced strata (Figure \ref{fig:SR_applied}).  This also is reflected by the number of strata with potential issues in the manual stratification issue table below. For example, stratum 1 below (white males) contains 3,314 patients, while stratum 3 (Native American males) contains only 18 patients, only one of whom was assigned to a surgical team.  In exceedingly large strata, matching becomes increasingly computationally intensive, while in exceedingly small and/or highly imbalanced strata, finding high quality matches can be infeasible (see section \ref{subsec:advanced_matching}).

\begin{example}
 ICU_mstrat$issue_table
\end{example}
\begin{example}
# A tibble: 16 x 6
   Stratum Treat Control Total Control_Proportion Potential_Issues                           
     <int> <int>   <int> <int>              <dbl> <chr>                                      
 1       1   761    2553  3314              0.770 none                          
 2       2   212     672   884              0.760 none                          
 3       3     1      17    18              0.944 Too few samples; Not en...
 4       4    13      67    80              0.838 Not enough treated samples    
 5       5    56     205   261              0.785 none                          
 6       6    65     286   351              0.815 Not enough treated samples    
 7       7    29     226   255              0.886 Not enough treated samples    
 8       8   174     563   737              0.764 none                          
 9       9   508    1842  2350              0.784 none                          
10      10   158     470   628              0.748 none                          
11      11     4      13    17              0.765 Too few samples               
12      12    15      54    69              0.783 Too few samples               
13      13    37     194   231              0.840 Not enough treated samples    
14      14    46     195   241              0.809 Not enough treated samples    
15      15    16     173   189              0.915 Not enough treated samples    
16      16   131     401   532              0.754 none   
\end{example}

\begin{figure}[t!]
\centering
\includegraphics[width = 2.75in]{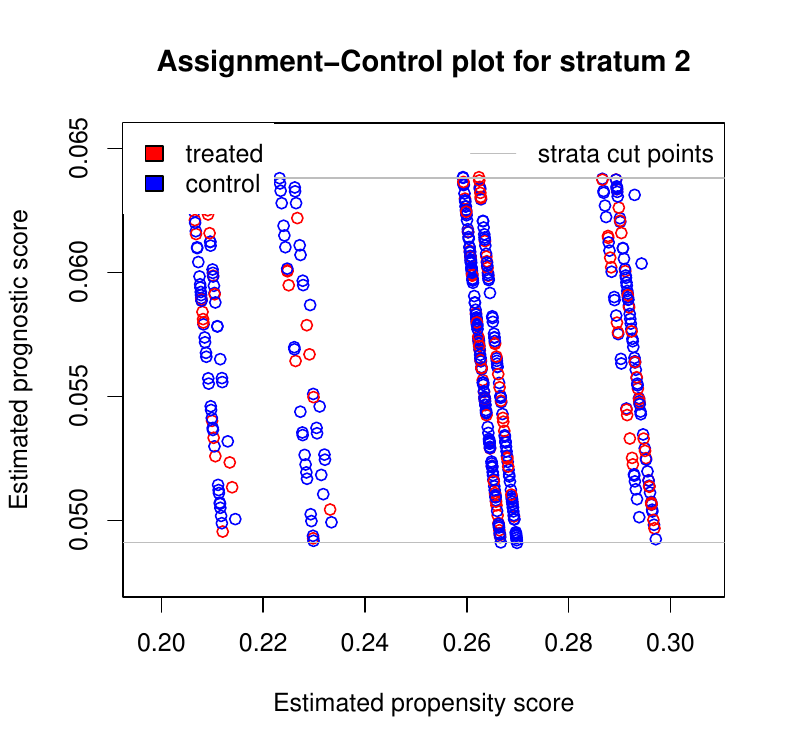}
\caption{\label{fig:AC_applied} Assignment-control plot for automatic stratification of ICU data.  The vertical striations are caused by heavily weighted discrete features in the propensity model, which cause points to align together.}
\end{figure}

The code below displays the assignment-control plot for one of the strata in the automatically stratified data set (Figure \ref{fig:AC_applied}). 

\begin{example}
plot(ICU_astrat, type = "AC",
    propensity = surgicalTeam ~ Female.pre + Birth.preTimeDays +
      RaceAsian.pre + RaceUnknown.pre + RaceOther.pre + RaceBlack.pre +
      RacePacificIslander.pre + RaceNativeAmerican.pre + all_latinos,
    stratum = 2)
\end{example}

The striae in this assignment-control plot appear when discrete characteristics (e.g. sex and race/ethnicity) are highly weighted in the propensity or prognostic score, causing observations to cluster together.  Since this is relatively common, \samp{jitter} arguments can be used to add small amounts of random noise to the coordinates of each point in order to avoid stacking.

\subsubsection{Matching}\label{subsubsec:real_matching}

After a suitable stratification is selected, observations can be matched within strata using \code{strata\_match}.  Since every stratum from the automatic stratification in this example contains at least a 1:2 ratio of patients who were assigned to surgical teams and those who were not, we can match 2 ``control'' (i.e., non-surgical team) patients to each ``treated'' (i.e., surgical team) subject in each stratum.  In this step, we match individuals who, based on their baseline covariates, appear equally likely to have been assigned to a surgical team vs. not.  
The following performs the matching:

\begin{example}
ICU_match <- strata_match(ICU_astrat, 
    propensity = surgicalTeam ~ Birth.preTimeDays + Female.pre +
      RaceAsian.pre + RaceUnknown.pre + RaceOther.pre + RaceBlack.pre + 
      RacePacificIslander.pre + RaceNativeAmerican.pre + all_latinos,
    k = 2)
\end{example}
\begin{example}
Fitting propensity model: surgicalTeam ~ Birth.preTimeDays + Female.pre +
RaceAsian.pre + RaceUnknown.pre + RaceOther.pre + RaceBlack.pre +
RacePacificIslander.pre + RaceNativeAmerican.pre + all_latinos +
strata(stratum)
\end{example}

Below, we print a summary:

\begin{example}
summary(ICU_match) 
\end{example}

\begin{example}
Structure of matched sets:
 1:2  0:1 
2226 2686 
Effective Sample Size:  2968 
(equivalent number of matched pairs).
\end{example}

At this point, the researcher can compare matched treated and control individuals to infer whether patients assigned to surgical treatment teams are more or less likely to be assigned a DNR code status, following up with sensitivity analyses (see, for example \CRANpkg{sensitivitymv} \citep{rosenbaum2018senmv})


\section{Key design choices and advanced functionality}\label{sec:key_design_choices}

\subsection{The selection of the pilot set}\label{subsec:pilot_sampling}

The previous illustrations demonstrated the simplest method of extracting the pilot set: a random subsampling of all controls.  \citet{aikens2020pilot} contains a more thorough discussion of the considerations that might inform the selection of a pilot set.

A first consideration is the pilot set size.  In general, the researcher should create a pilot set large enough to build  a reliable prognostic model and retain enough remaining controls to select high-quality matches to the treatment group.  This  depends on the quality and number of available controls and the relative difficultly of fitting a prognostic model on the measured covariates. When high-quality controls (i.e. those resembling the treatment group) are scarce, the researcher should consider a smaller pilot set or a different study design altogether.

Another consideration is composition. Ideally, the individuals in the pilot set should be similar to the individuals in the treatment group, so a prognostic model built on this pilot set will not be extrapolating heavily when estimating prognostic scores on the analysis set. This approach can be especially important when there is some category of observations in the data which are relatively rare, and the researcher would like to ensure that some observations in this category end up in both the pilot and analysis sets. When discrete covariates are specified with the \samp{group\_by\_covariates} argument to \code{auto\_stratify} the pilot set will be split proportionally based on these covariates, so that the pilot set will be representative of the total control sample in terms of these covariates. This option can be used directly with \code{auto\_stratify}. However, the \code{split\_pilot\_set} function is supplied as a convenience for users who prefer to split the pilot set themselves before stratification, as demonstrated below.

\begin{example}
ICU_split <- split_pilot_set(ICU_data, treat = "surgicalTeam", 
    pilot_fraction = 0.1, group_by_covariates = c("Female.pre", "self_pay")) 
\end{example}
\begin{example}
Constructing a pilot set by subsampling 10
Subsampling while balancing on:
Female.pre self_pay 
\end{example}

\code{ICU\_split}, above, is a list containing a \code{pilot\_set} and an \code{analysis\_set}, partitioned while balancing sex and payment method (i.e. insurance or self-pay). Once this is done, the results can be passed to \code{auto\_stratify} such as with the code below:

\begin{example}
ICU_astrat2 <- auto_stratify(data = ICU_split$analysis_set,
    treat = "surgicalTeam", 
    prognosis = DNR ~ Birth.preTimeDays + Female.pre + RaceAsian.pre + 
      RaceUnknown.pre + RaceOther.pre + RacePacificIslander.pre + 
      RaceBlack.pre + RaceNativeAmerican.pre + all_latinos,
    pilot_sample = ICU_split$pilot_set, size = 500)
\end{example}

\subsection{Fitting the prognostic model}\label{subsec:model_fitting}

To fit the prognostic model, \code{auto\_stratify} uses either linear (continuous outcome) or logistic regression (binary outcome). To accommodate a wider variety of modeling choices, \code{auto\_stratify} can also be run using a vector of analysis set prognostic scores or prognostic model object\footnote{Model objects must have a method associated with the \code{predict} generic function}.

The example below uses the \CRANpkg{glmnet} package \citep{friedman2010regularization} to fit a cross-validated lasso on the pilot set which was extracted in the previous section.

\begin{example}
library("glmnet")
x_pilot <- ICU_split$pilot_set 
    dplyr::select(Birth.preTimeDays, Female.pre, RaceAsian.pre,
      RaceUnknown.pre, RaceOther.pre, RaceBlack.pre, 
      RacePacificIslander.pre, RaceNativeAmerican.pre, all_latinos) 
    as.matrix()
y_pilot <- ICU_split$pilot_set 
    dplyr::select(DNR) 
    as.matrix()

cvfit <- cv.glmnet(x_pilot, y_pilot, family = "binomial")
\end{example}

The prognostic scores can then be estimated on the analysis set:

\begin{example}
x_analysis <- ICU_split$analysis_set 
    dplyr::select(Birth.preTimeDays, Female.pre, RaceAsian.pre,
      RaceUnknown.pre, RaceOther.pre, RaceBlack.pre, 
      RacePacificIslander.pre, RaceNativeAmerican.pre, all_latinos) 
    as.matrix()

lasso_scores <- predict(cvfit, newx = x_analysis, s = "lambda.min",
    type = "response")
\end{example}

Finally, these scores can be passed to \code{auto\_stratify} with the \samp{prognosis} argument, producing a stratified data set which can be examined further with \pkg{stratamatch} diagnostic tools.

\begin{example}
ICU_astrat3 <- auto_stratify(data = ICU_split$analysis_set,
    treat = "surgicalTeam", outcome = "DNR", prognosis = lasso_scores,
    pilot_sample = ICU_split$pilot_set, size = 500)
\end{example}

\subsection{Matching}\label{subsec:advanced_matching}

Section \ref{sec:illustrations} demonstrates how the \pkg{stratamatch} package can be used for optimal $1:k$ matching on propensity score.  If desired, a data set stratified with \pkg{stratamatch} can instead be matched within strata using other matching software (e.g., \pkg{optmatch} \citep{hansen2006optmatch}, or \pkg{matchit} \citep{ho2011matchit}).  For example, users proficient with \pkg{optmatch} will note that adding \code{+ strata(stratum)} to the matching formula supplied to \code{optmatch::pairmatch} and other matching functions will match within stratum assignments in the analysis set. 

More nuanced matching schemes may also help address imbalances in the number of treated and control units within strata.  For example, the researcher could perform $1:k$ matching within each stratum, but allow $k$ to vary between strata - matching more controls to each treated individual in strata where controls are plentiful and performing $1:1$ or $1:2$ matching where controls are less abundant.  Another solution is to use a matching scheme within strata which naturally allows for variation in the ratio of treated and control individuals in matched sets, such as fullmatching \citep{rosenbaum1991fullmatch} or variable k matching \citep{pimentel2015variable}.

\begin{figure}[htbp]
\centering
    \includegraphics[width = 2.5in]{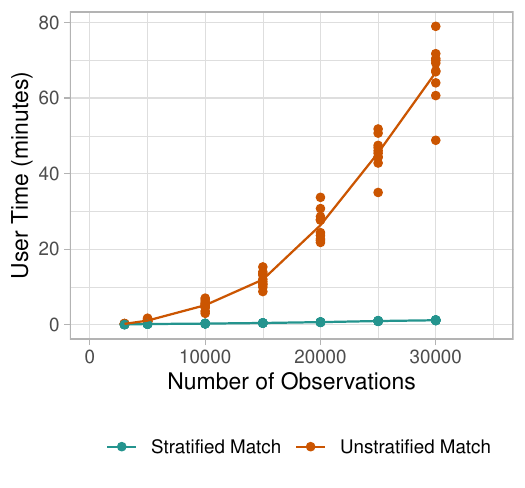}
    \caption{Measured computation times for stratified and unstratified matching on a modern laptop. Unstratified matching scales in a supra-linear manner with sample size \cite{hansen2006optmatch}, while stratified matching with a set strata size tends to scale more favorably with $n$. At sample sizes of  30,000, optimally matching a whole data set may take over an hour, and much larger sample sizes may quickly become infeasible.}
    \label{fig:performance}
\end{figure}

As shown in figure \ref{fig:performance}, stratification is expected to substantially accelerate the matching process, especially for large sample sizes (several thousand or more). Hansen and Klopfer articulate a worst-case run-time for various forms of optimal matching with \pkg{optmatch} as $O(n^3\log(nM))$, where $M$ represents the maximum matching discrepancy between treated and control observations \citep{hansen2006optmatch}. For context, this scales slightly less favorably than matrix inversion, which quickly becomes time-consuming for large inputs.  By comparison, matching within strata of a fixed size tends to scale much more favorably for large $n$ (figure \ref{fig:performance}). To further accelerate computation, a researcher might distribute matching the stratified data set over several computing nodes.

\section{Trouble-shooting a stratification scheme} \label{sec:troubleshooting}

This section summarizes some common pitfalls and workarounds while stratifying a data set.  Importantly, in order to preserve the separation of the design and analysis set, individuals partitioned into the pilot set must not be recombined with the analysis set.  For instance, simply running \code{auto\_stratify} repeatedly with different seeds to sample new pilot sets from the data and fit new prognostic score models may lead to overfitting of the prognostic model, raising concerns of bias in the study results (see \citet{hansen2008prognostic, abadie2018endogenous}).

The following issues are common:
\begin{enumerate}
    \item \textbf{Some strata are too small or too large:} Often, this problem can be solved simply by rerunning \code{auto\_stratify} with a different \samp{size} parameter.  When this is done, the researcher should be sure to use the same pilot and analysis set as they received when they first ran \code{auto\_stratify} (i.e., do not partition a new pilot set).
    \item \textbf{The strata have poor balance of treated and control individuals:} This situation is relatively common, but often straightforward to address with matching schemes that match more controls to each treated observation or  allow for variable treat:control ratios. See section \ref{subsec:advanced_matching} for some suggestions.
    \item \textbf{The prognostic model is poor:}
    In some cases, the user may encounter an error fitting the prognostic model, or they may suspect from prognostic model diagnostics that the model does a poor job of capturing variation predictive of the outcome.  There are a few reasons the prognostic model may be problematic.
    \begin{enumerate}
        \item \textit{The prognostic model was mis-specified.} In this case, the user should fit a revised prognostic model on the same pilot set as was previously used. However, refitting repeatedly can lead to overfitting, so this should be done in moderation.
        \item \textit{The pilot set was too small to get a reliable fit.} In this case, the user can add more samples from the analysis set to the existing pilot set.  Samples that are moved into the pilot set must stay in the pilot set and should not be re-pooled with the analysis set.
        \item \textit{Pilot set size is sufficient, but prognostic model perfectly separates treated individuals from control individuals:} If this occurs in either the pilot set or analysis set, it may be a sign that overlap is poor.  See below.
    \end{enumerate}
    \item \textbf{The treated and control individuals within strata have poor overlap in propensity and/or prognostic scores:} This problem is best diagnosed with assignment-control plots (see \citet{aikens2020pilot} for a deeper description). Propensity and prognostic score based subclassification methods both depend on some form of overlap in the baseline characteristics of treated and control individuals in order to make a valid estimate of causal effect (for a summary, see \citet{leacy2014joint}).  Treatment and control groups which are clearly separated in terms of either their propensity scores or prognostic scores can be an indication that these two groups should not be compared, because the resulting inference on treatment effect would be misleading.  A researcher facing this situation might consider trimming the score space \citep{glynn2019comparison} in some cases, or seeking out another data set if the overlap problems are severe. While this may seem to be a disappointing result, the ability to identify these data issues before proceeding is one of the most important strengths of design-based causal inference (see, for example \cite{austin2011introduction}).
\end{enumerate}


\section{Summary and discussion} \label{sec:discussion}
Stratifying a data set prior to matching may make optimal and fullmatching designs scale more practically for modern observational sample sizes. However, the primary objective of \pkg{stratamatch} is not to directly implement a computationally taxing task, but to expand access to sophisticated study design tools for a wide range of researchers with varying levels of technical and statistical sophistication.  Indeed, the computational steps of stratification are relatively straightforward; however, the statistical concept of the pilot design is nuanced, and the process of stratifying a data set and interrogating the quality of that stratification can be thought-intensive and isn't well-supported by other resources. The \pkg{stratamatch} package is intended to make a prognostic score stratification pilot designs -- and stratified matching designs in general -- easily implementable, with helpful diagnostic tools and documentation. The overall goal of this effort is to push researchers toward approaches and diagnostics which emphasize stronger study design in the observational setting.  In modern observational studies, designs such as the \pkg{stratamatch} approach which are \textit{tailored} to large-sample studies can offer increased precision and other statistical benefits that might otherwise be left on the table by more traditional approaches.

\bibliography{aikens_et_al}

\begin{thebibliography}{26}
\providecommand{\natexlab}[1]{#1}
\providecommand{\url}[1]{\texttt{#1}}
\expandafter\ifx\csname urlstyle\endcsname\relax
  \providecommand{\doi}[1]{doi: #1}\else
  \providecommand{\doi}{doi: \begingroup \urlstyle{rm}\Url}\fi

\bibitem[Abadie et~al.(2018)Abadie, Chingos, and West]{abadie2018endogenous}
A.~Abadie, M.~M. Chingos, and M.~R. West.
\newblock Endogenous stratification in randomized experiments.
\newblock \emph{Review of Economics and Statistics}, 100\penalty0 (4):\penalty0
  567--580, 2018.

\bibitem[Aikens et~al.(2020)Aikens, Greaves, and Baiocchi]{aikens2020pilot}
R.~C. Aikens, D.~Greaves, and M.~Baiocchi.
\newblock A pilot design for observational studies: Using abundant data
  thoughtfully.
\newblock \emph{Statistics in Medicine}, 39\penalty0 (30):\penalty0 4821--4840,
  2020.

\bibitem[Antonelli et~al.(2018)Antonelli, Cefalu, Palmer, and
  Agniel]{antonelli2018doubly}
J.~Antonelli, M.~Cefalu, N.~Palmer, and D.~Agniel.
\newblock Doubly robust matching estimators for high dimensional confounding
  adjustment.
\newblock \emph{Biometrics}, 74\penalty0 (4):\penalty0 1171--1179, 2018.

\bibitem[Austin(2011)]{austin2011introduction}
P.~C. Austin.
\newblock An introduction to propensity score methods for reducing the effects
  of confounding in observational studies.
\newblock \emph{Multivariate behavioral research}, 46\penalty0 (3):\penalty0
  399--424, 2011.

\bibitem[Bertsekas and Tseng(1988)]{RELAX1988}
D.~P. Bertsekas and P.~Tseng.
\newblock Relaxation methods for minimum cost ordinary and generalized network
  flow problems.
\newblock \emph{Operations Research}, 36\penalty0 (1):\penalty0 93--114, 1988.

\bibitem[Chavez et~al.(2018)Chavez, Richman, Kaimal, Bentley, Yasukawa, Altman,
  Periyakoil, and Chen]{chavez2018reversals}
G.~Chavez, I.~B. Richman, R.~Kaimal, J.~Bentley, L.~A. Yasukawa, R.~B. Altman,
  V.~S. Periyakoil, and J.~H. Chen.
\newblock Reversals and limitations on high-intensity, life-sustaining
  treatments.
\newblock \emph{PloS one}, 13\penalty0 (2):\penalty0 e0190569, 2018.

\bibitem[Friedman et~al.(2010)Friedman, Hastie, and
  Tibshirani]{friedman2010regularization}
J.~Friedman, T.~Hastie, and R.~Tibshirani.
\newblock Regularization paths for generalized linear models via coordinate
  descent.
\newblock \emph{Journal of Statistical Software}, 33\penalty0 (1):\penalty0
  1--22, 2010.
\newblock URL \url{http://www.jstatsoft.org/v33/i01/}.

\bibitem[Glynn et~al.(2019)Glynn, Lunt, Rothman, Poole, Schneeweiss, and
  St{\"u}rmer]{glynn2019comparison}
R.~J. Glynn, M.~Lunt, K.~J. Rothman, C.~Poole, S.~Schneeweiss, and
  T.~St{\"u}rmer.
\newblock Comparison of alternative approaches to trim subjects in the tails of
  the propensity score distribution.
\newblock \emph{Pharmacoepidemiology and drug safety}, 28\penalty0
  (10):\penalty0 1290--1298, 2019.

\bibitem[Goodman et~al.(2017)Goodman, Schneeweiss, and
  Baiocchi]{goodman2017designthinking}
S.~N. Goodman, S.~Schneeweiss, and M.~Baiocchi.
\newblock Using design thinking to differentiate useful from misleading
  evidence in observational research.
\newblock \emph{Jama}, 317\penalty0 (7):\penalty0 705--707, 2017.

\bibitem[Hansen(2008)]{hansen2008prognostic}
B.~B. Hansen.
\newblock The prognostic analogue of the propensity score.
\newblock \emph{Biometrika}, 95\penalty0 (2):\penalty0 481--488, 2008.

\bibitem[Hansen and Klopfer(2006)]{hansen2006optmatch}
B.~B. Hansen and S.~O. Klopfer.
\newblock Optimal full matching and related designs via network flows.
\newblock \emph{Journal of Computational and Graphical Statistics}, 15\penalty0
  (3):\penalty0 609--627, 2006.

\bibitem[Hern{\'a}n and Robins(2016)]{hernan2016using}
M.~A. Hern{\'a}n and J.~M. Robins.
\newblock Using big data to emulate a target trial when a randomized trial is
  not available.
\newblock \emph{American journal of epidemiology}, 183\penalty0 (8):\penalty0
  758--764, 2016.

\bibitem[Ho et~al.(2011)Ho, Imai, King, Stuart, et~al.]{ho2011matchit}
D.~E. Ho, K.~Imai, G.~King, E.~A. Stuart, et~al.
\newblock \pkg{MatchIt}: Nonparametric preprocessing for parametric causal
  inference.
\newblock \emph{Journal of Statistical Software, http://gking. harvard.
  edu/matchit}, 2011.

\bibitem[King and Nielsen(2016)]{king2016propensity}
G.~King and R.~Nielsen.
\newblock Why propensity scores should not be used for matching.
\newblock \emph{Copy at http://j. mp/1sexgVw Download Citation BibTex Tagged
  XML Download Paper}, 378, 2016.

\bibitem[Leacy and Stuart(2014)]{leacy2014joint}
F.~P. Leacy and E.~A. Stuart.
\newblock On the joint use of propensity and prognostic scores in estimation of
  the average treatment effect on the treated: A simulation study.
\newblock \emph{Statistics in medicine}, 33\penalty0 (20):\penalty0 3488--3508,
  2014.

\bibitem[Lohr(2019)]{lohr2019sampling}
S.~L. Lohr.
\newblock \emph{Sampling: Design and analysis: Design and analysis}.
\newblock CRC Press, 2019.

\bibitem[Pimentel et~al.(2015)Pimentel, Yoon, and Keele]{pimentel2015variable}
S.~D. Pimentel, F.~Yoon, and L.~Keele.
\newblock Variable-ratio matching with fine balance in a study of the peer
  health exchange.
\newblock \emph{Statistics in medicine}, 34\penalty0 (30):\penalty0 4070--4082,
  2015.

\bibitem[Rigdon et~al.(2018)Rigdon, Baiocchi, and Basu]{rigdon2018nearfar}
J.~Rigdon, M.~Baiocchi, and S.~Basu.
\newblock Near-far matching in {R}: The {nearfar} package.
\newblock \emph{Journal of Statistical Software, Code Snippets}, 86\penalty0
  (5):\penalty0 1--21, 2018.
\newblock \doi{10.18637/jss.v086.c05}.

\bibitem[Rosenbaum(1991)]{rosenbaum1991fullmatch}
P.~R. Rosenbaum.
\newblock A characterization of optimal designs for observational studies.
\newblock \emph{Journal of the Royal Statistical Society B}, 53\penalty0
  (3):\penalty0 597--610, 1991.

\bibitem[Rosenbaum(2005{\natexlab{a}})]{rosenbaum2005heterogeneity}
P.~R. Rosenbaum.
\newblock Heterogeneity and causality: Unit heterogeneity and design
  sensitivity in observational studies.
\newblock \emph{The American Statistician}, 59\penalty0 (2):\penalty0 147--152,
  2005{\natexlab{a}}.

\bibitem[Rosenbaum(2005{\natexlab{b}})]{rosenbaum2005sensitivity}
P.~R. Rosenbaum.
\newblock Sensitivity analysis in observational studies.
\newblock \emph{Encyclopedia of statistics in behavioral science}, 4:\penalty0
  1809--1814, 2005{\natexlab{b}}.

\bibitem[Rosenbaum(2018)]{rosenbaum2018senmv}
P.~R. Rosenbaum.
\newblock \emph{sensitivitymv: Sensitivity Analysis in Observational Studies},
  2018.
\newblock URL \url{https://CRAN.R-project.org/package=sensitivitymv}.
\newblock R package version 1.4.3.

\bibitem[Rosenbaum(2019)]{rosenbaum2019DOS2}
P.~R. Rosenbaum.
\newblock \emph{DOS2: Design of Observational Studies, Companion to the Second
  Edition}, 2019.
\newblock URL \url{https://CRAN.R-project.org/package=DOS2}.
\newblock R package version 0.5.2.

\bibitem[Rosenbaum and Rubin(1983)]{rosenbaum1983central}
P.~R. Rosenbaum and D.~B. Rubin.
\newblock The central role of the propensity score in observational studies for
  causal effects.
\newblock \emph{Biometrika}, 70\penalty0 (1):\penalty0 41--55, 1983.

\bibitem[Rosenbaum et~al.(2010)]{rosenbaum2010designbook}
P.~R. Rosenbaum et~al.
\newblock \emph{Design of Observational Studies}, volume~10.
\newblock Springer-Verlag, 2010.

\bibitem[Rubin et~al.(2008)]{rubin2008design}
D.~B. Rubin et~al.
\newblock For objective causal inference, design trumps analysis.
\newblock \emph{The Annals of Applied Statistics}, 2\penalty0 (3):\penalty0
  808--840, 2008.

\end{thebibliography}

\address{Rachael C. Aikens\\
  Interdepartmental Program in Biomedical Informatics\\
  Stanford University\\
  Stanford, CA\\
  USA}

\address{Joseph Rigdon\\
  Department of Biostatistics and Data Science\\
  Wake Forest School of Medicine\\
  Winston-Salem, North Carolina}

\address{Justin Lee\\
  Quantitative Sciences Unit\\
  Stanford University\\
  Stanford, CA}

\address{Michael Baiocchi\\
  Epidemiology and Population Health\\
  Stanford University\\
  Stanford, CA}

\address{Andrew B. Goldstone\\
  Division of Cardiovascular Surgery\\
  University of Pennsylvania\\
  Philadelphia, PA}
  
\address{Peter Chiu\\
  Department of Cardiothoracic Surgery\\
  Stanford University School of Medicine\\
  Stanford, CA}

\address{Y. Joseph Woo\\
  Department of Cardiothoracic Surgery \\
  Stanford University School of Medicine\\
  Stanford, CA}
  
\address{Jonathan H. Chen\\
  Biomedical Informatics Research Institute \\
  Stanford University Medical Center\\
  Stanford, CA \\
  \email{jonc101@stanford.edu}}
\end{article}

\end{document}